# Structural Vulnerability Assessment in Urban Transport Networks: A Network-Wide Geometric Approach Using Gromov–Wasserstein

**Iman Seyedi[1,*], Antonio Candelieri[2], Enza Messina[1], and Francesco Archetti[1]**

[1] University of Milano-Bicocca, Department of Computer Science Systems and Communication, Italy; seyediman.seyedi@unimib.it, enza.messina@unimib.it, and francesco.archetti@unimib.it

[2] University of Milano-Bicocca, Department of Economics Management and Statistics, Italy; antonio.candelieri@unimib.it

* Correspondence: seyediman.seyedi@unimib.it

**Abstract:** Urban transportation networks are inherently vulnerable to disruptions that affect connectivity and passenger mobility. Traditional graph-theoretic metrics, such as betweenness and degree centrality, offer insights into *local* network structure but often fail to capture global structural distortions resulting from link failures. On the other hand, global indices, such as those based on spectral analysis of the network's graph, fail in identifying critical elements. This study proposes to quantify the structural modifications implied by the disruption of single elements in a transportation network through the Gromov–Wasserstein distance. Specifically, we iteratively remove one single edge from the original network to simulate a disruptive event and then compute the Gromov-Wasserstein distance between the original network and the "disrupted" one. Finally, edges are ranked depending on the observed Gromov-Wasserstein distance: the higher the value of the distance, the more critical the edge is in terms. Two transportation networks from Berlin are considered in the experiments, namely Berlin Friedrichshain Center (BFC) and Berlin Tiergarten (BT). Results reveal that Gromov-Wasserstein is largely uncorrelated with edge betweenness ($\rho < 0.1$), proving its ability to capture vulnerability aspects overlooked by local network measures. Moreover, Gromov-Wasserstein exhibits an almost perfect correlation ($\rho \approx 0.9999$) against a proxy measure of the transportation service level, that is, the increase in the maximum shortest path. As a result, the Gromov-Wasserstein distance can be used to rank edges depending on their criticality with respect to their individual impact on the overall infrastructure and level, allowing for prioritizing maintenance, emergency planning, and enhancing the resilience of the urban transport network.

**Keywords:** Urban transport networks; Network resilience; Gromov–Wasserstein distance; Graph theory; Maximum shortest path.

## 1. Introduction

Urban transport networks are critical infrastructures that underpin the daily functioning of cities. They provide the backbone for mobility, emergency response, and economic activity. Understanding their resilience – the ability to maintain operational functionality under failures or disruptions – is essential for effective urban planning, disaster management, and sustainable mobility ([1], [2]). Disruptions may arise from accidents, construction, natural hazards, or deliberate attacks, significantly affecting accessibility and efficiency. Traditionally, graph-theoretic measures have been employed to evaluate network resilience.

Metrics such as node degree, edge betweenness centrality, closeness, and algebraic connectivity quantify structural properties associated with vulnerability [3]. For example, betweenness centrality measures how frequently an edge or node lies on the shortest paths between pairs of nodes, offering insights into potential bottlenecks. However, these measures primarily focus on local or path-based properties, overlooking the global geometry and distribution of shortest paths across the network [4]. Consequently, edges that are structurally critical from a global perspective may be missed by classical metrics. This limitation has motivated researchers to seek alternative approaches that can capture the holistic geometric properties of networks. The growing literature on transport vulnerability studies has increasingly recognized the need for methods that go beyond topological properties to represent the demand and spatial characteristics of network systems [5], [6]. In this context, the concept of measuring structural distortions across entire network geometries has gained attention, particularly for spatially embedded networks where the spatial arrangement of nodes and the distribution of shortest paths matter as much as individual connectivity patterns. The challenge lies in developing metrics that are sensitive to outliers and can capture the overall structural reorganization of networks under perturbations[7].

Recent developments in optimal transport theory provide new tools for evaluating network vulnerability. The classical Wasserstein distance, originally developed for comparing probability distributions, has found applications in network analysis by treating node distributions and connectivity patterns as transport problems [8]. However, when applied to network comparison, the standard Wasserstein distance faces a fundamental limitation: it requires networks to be embedded in the same metric space and assumes a fixed correspondence between nodes across different network states [7]. In the context of network resilience analysis, where edge removals fundamentally alter the underlying metric structure and shortest-path distances, this assumption becomes problematic. Specifically, the classical Wasserstein distance cannot adequately capture how the removal of critical edges reshapes the entire geometric relationship between nodes, as it lacks the flexibility to compare networks with inherently different metric structures. The Gromov–Wasserstein (GW) distance extends the classical Wasserstein distance to compare entire metric spaces, making it particularly suitable for networks with spatially embedded nodes [7]. In urban road networks, the GW distance quantifies the overall structural distortion induced by the removal of an edge, capturing not only changes in individual shortest paths but also how the entire network geometry reorganizes. This provides a global perspective on resilience, bridging the gap between local flow-based metrics and large-scale structural vulnerability.

Among probabilistic approaches, early work by [9]. introduced shortest-path based comparisons coupled with symmetrized Kullback–Leibler divergence, illustrating how distributional perspectives extend beyond single metrics. More generally, many divergences and distances have been proposed, but most remain insufficient for capturing the full network geometry.

Building on this line of probabilistic and geometry-aware methods, Optimal Transport (OT) has emerged as a principled framework for network analysis and quantifying structural dissimilarity. The classical Wasserstein distance provides global structural measures by quantifying transport costs across probability distributions [10], [11]. Unlike simple pointwise comparisons, it lifts entire distributions into a geometry-aware metric space, offering continuity, convexity, and—in some cases—differentiability, which makes it attractive for optimization and learning tasks. However, a key limitation of the Wasserstein distance is that it requires both measures to reside in the same metric space. This assumption is often violated in network vulnerability assessment, where edge removals alter the network's intrinsic geometry. The GW distance overcomes this by comparing internal distance structures, without requiring a shared embedding [12], [13].

GW distance is particularly well-suited for resilience analysis, as it captures both localized disruptions and their global propagation across the network, integrating aspects that classical centrality-based metrics may overlook. While OT methods have been extensively studied in machine learning and graph matching [14], [15], [16], their application to network vulnerability assessment remains limited. Leveraging GW distance thus represents a significant opportunity to advance resilience analysis by providing a holistic measure that unifies extreme-case disruptions with distributed structural changes.

This study investigates the resilience of the transportation network of two Berlin neighborhoods – Berlin Friedrichshain Center (BFC) and Berlin Tiergarten (BT) – by analyzing the effect of edge removals from the network's infrastructure through the GW distance. Baseline distance matrices representing shortest-path distances between all nodes in each network are constructed, and edge removal experiments are conducted to calculate GW distances between the baseline and perturbed networks. To validate the effectiveness of GW distance and understand what aspects of network vulnerability it captures, we compare it against a widely adopted local measure, namely the edge betweenness, and a *proxy* of the change in the level of service, that is, the difference in terms of maximum shortest path (aka network's diameter) implied by the removal of an edge.

Our comparative analysis reveals that GW distance exhibits an almost perfect correlation with, indicating that it effectively captures extreme-case connectivity deterioration while simultaneously accounting for the entire distribution of shortest-path changes across the network. This dual sensitivity—to both worst-case scenarios and distributed structural shifts—distinguishes GW from classical metrics such as edge betweenness centrality, which showed very low correlation with both GW and ΔMSP. These findings demonstrate that GW distance encodes a broader range of structural distortions than traditional flow-based measures, capturing network features that extend beyond local bottleneck identification. The study identifies the most critical edges in each network based on GW distance rankings, providing insights for resilience planning and infrastructure prioritization.

The remainder of the paper is organized as follows. Section 2 reviews related work on network resilience, classical graph metrics, and the application of optimal transport methods to networks. Section 3 describes the datasets and preprocessing steps, including network statistics and the spatial embedding of nodes. Section 4 presents the methodology, detailing the construction of distance matrices, the computation of Gromov–Wasserstein distances, edge removal experiments, and Maximum Shortest Path (MSP) analysis. Section 5 reports the results, including the identification of critical edges, correlation analyses, and visualizations of network impacts. Finally, Section 6 discusses the implications of the findings, highlights the advantages of GW distance over traditional measures such as edge betweenness, and suggests potential directions for future research.

## 2. Related Works

Public transportation systems play a critical role in urban mobility by reducing traffic congestion and providing environmentally sustainable alternatives for commuting [17]. Transport network vulnerability refers to the system's ability to maintain service under adverse conditions, typically quantified as connectivity deterioration or increases in travel times resulting from infrastructure disruptions [18], [19]. More broadly, it is essential to emphasize that networks—whether engineered systems, such as transportation, gas, and water infrastructures [20], or naturally occurring systems, like ecological food webs [21] – can all be represented as graphs. By taking a graph-theoretic or topological (no-flow) view, one can already extract significant information about the underlying structure and its vulnerabilities, even before introducing dynamic flow or demand models. In this perspective, resilience to natural (e.g., earthquakes, tsunamis) or man-made (e.g., deliberate attacks, failures) events can be captured as a change in the network structure [22]. A key principle is that resilience is associated with dissimilarity between graphs: if a disrupted network diverges strongly from its baseline structure, resilience is low [23]. Various notions of graph dissimilarity exist, including isomorphism-based measures, centrality-driven metrics such as betweenness, or distance-based divergences [24].

Classical network analyses often rely on graph-theory metrics, including degree centrality, betweenness, closeness, and network diameter, to identify critical nodes or links and assess resilience under simulated disruptions [25], [26], [27]. While these measures provide valuable insights into local connectivity patterns and individual shortest paths, they primarily capture localized or extreme structural features and may overlook distributed, network-wide changes in connectivity that occur under perturbations [28], [29]. This issue is particularly pronounced when assessing edges that are structurally critical from a global perspective but may not exhibit high centrality scores. The limitations of classical graph-theoretic metrics in capturing global structural distortions have driven researchers toward more comprehensive approaches for network vulnerability assessment.

To address these gaps, researchers have explored several complementary approaches. Some studies have incorporated traveler behavior and operational attributes into vulnerability assessments, considering adaptive route choices, passenger welfare, and changes in travel times [28], [30]. Multilayer network theory has been applied to represent interactions among different transport modes, enabling more accurate modeling of cascading failures and passenger redistribution [31], [32], [33]. However, these approaches, while capturing dynamic flows and congestion effects, still rely fundamentally on topological metrics and therefore cannot account for holistic geometric changes arising from network perturbations. The recognition that network resilience requires methods sensitive to overall structural reorganization has led to increased interest in approaches that can measure geometric distortions across entire networks. For urban road networks, this challenge is particularly acute given their spatial embedding and the importance of shortest-path distributions for accessibility [34], [35]. Traditional GIS-based modeling and topology optimization provide spatial accuracy, but they remain focused on localized disruptions and often demand extensive preprocessing [2], [36], [37].

Existing studies on transport network vulnerability, using graph-theoretic metrics, multilayer models, or classical Wasserstein distance, often fail to capture global geometric changes caused by disruptions. Local metrics like betweenness centrality miss distributed structural shifts, while flow-based models require extensive preprocessing and remain topology-focused. The classical Wasserstein distance is limited by its need for a shared metric space, unsuitable for networks with altered geometries. Thus, there is a need for a holistic, geometry-aware metric to quantify network-wide structural distortions. This study addresses this gap by applying the Gromov–Wasserstein distance to assess urban transport network resilience.

## 3. Materials

### 3.1. Data description

We analyze two urban transportation networks in Berlin: the Berlin Friedrichshain Center (BFC) and the Berlin Tiergarten (BT), as downloaded from TransportationNetworks[1].

The main goal is to study network resilience and edge criticality by simulating edge removals and evaluating their impact on the overall network structure, following the framework described by [1]. The networks are represented as spatial undirected graphs $G = (V, E)$, where the nodes set $V$ consists of intersections or transit points, and the edges set $E$ consists of road segments connecting pairs of nodes. Each node has its own spatial coordinates $(lat_i, lon_i)$, enabling a bi-dimensional representation of the network. We assume that the cardinality of the nodes set is $|V| = n$ and cardinality of the edges set is $|E| = m$. For the BFC network we have $n = 224$ nodes and $m = 523$ edges, while for BT we have $n = 361$ nodes and $m = 766$ edges.

The bi-dimensional representations of two networks are reported in Figures 1 and 2: it is evident that the BFC network is the relatively dense urban core with short links and multiple intersecting paths, while the BT network is larger and exhibits longer road segments with fewer intersections, reflecting different urban layouts and functional roles in Berlin's transportation system. These visualizations provide an intuitive understanding of the network structures and highlight potential areas of vulnerability.

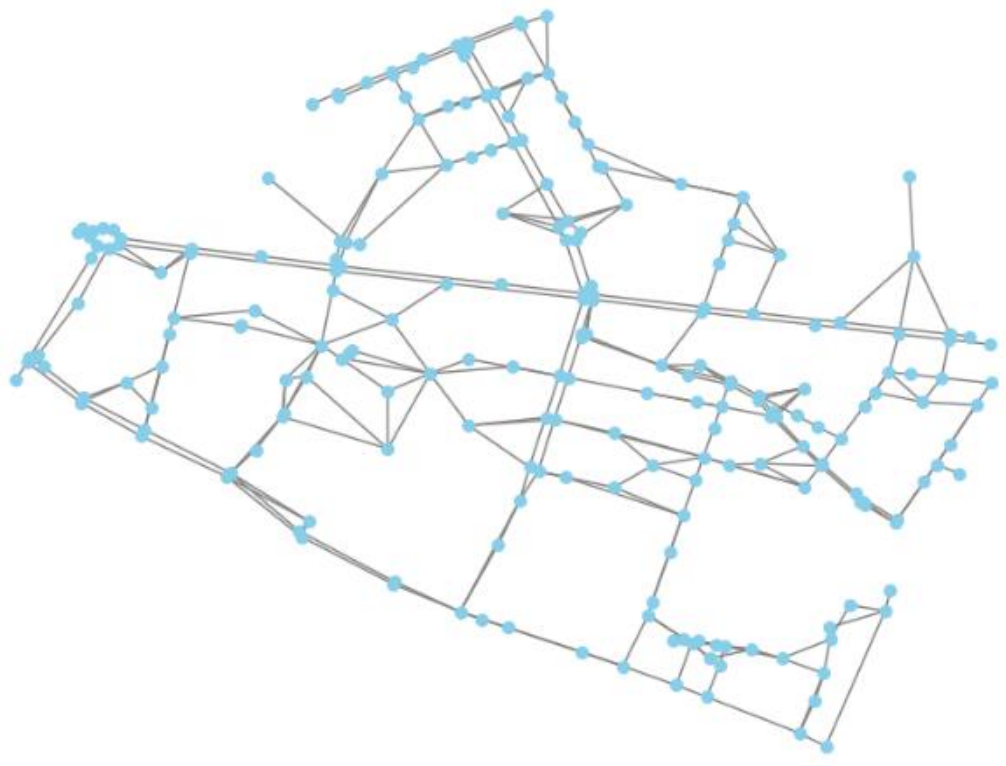

Figure. BFC Transportation Network

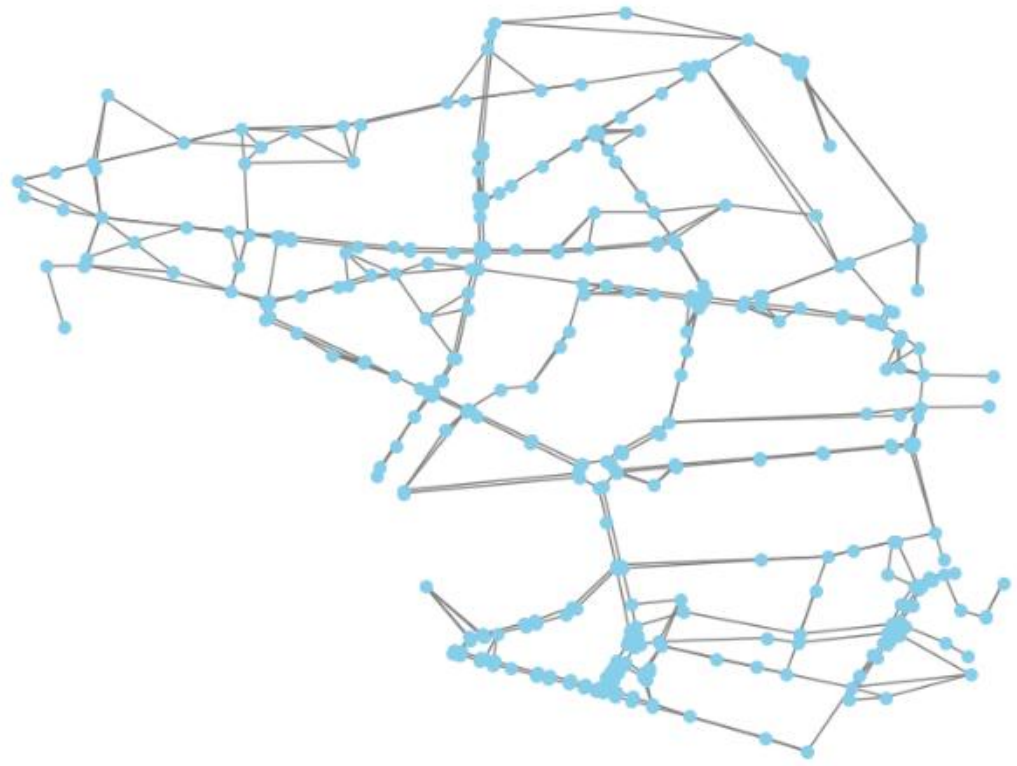

Figure 2. BT Transportation Network

### 3.2. Local graph-based measures and their distributions as global indicators

Table 1 presents the most widely used local graph-based measures of the two networks, including the number of nodes and edges, minimum and maximum node degree, average degree and variance, network density, and diameter. These metrics allow for a quantitative comparison of network topology and highlight key differences in structural robustness depending on the distribution of local measures referring to individual elements (i.e., edges). For instance, while the average degree is similar for BFC (3.357) and BT (3.3148), the degree variance is higher in BFC, indicating a more heterogeneous distribution of connectivity among nodes. The diameters (23 for BFC and 31 for BT) reflect the longest shortest paths in each network, giving an initial sense of worst-case accessibility across the network.

[1] https://github.com/bstabler/TransportationNetworks

**Table 1.** Network statistics for BFC and BT

| Network | Nodes | Edges | $min(d_k)$ | $max(d_k)$ | $\bar{d}$ | $\sigma^2(d_k)$ | Density | Diameter |
|---|---|---|---|---|---|---|---|---|
| BT | 361 | 766 | 1 | 9 | 3.3148 | 1.0937 | 0.00926 | 31 |
| BFC | 224 | 523 | 1 | 8 | 3.357 | 1.2737 | 0.01505 | 23 |

The degree distributions of the two networks are shown in Figures 3 and 4: the BFC's node degree distribution is relatively balanced with a few highly connected nodes acting as hubs, while the BT's node degree distribution is slightly more heterogeneous.

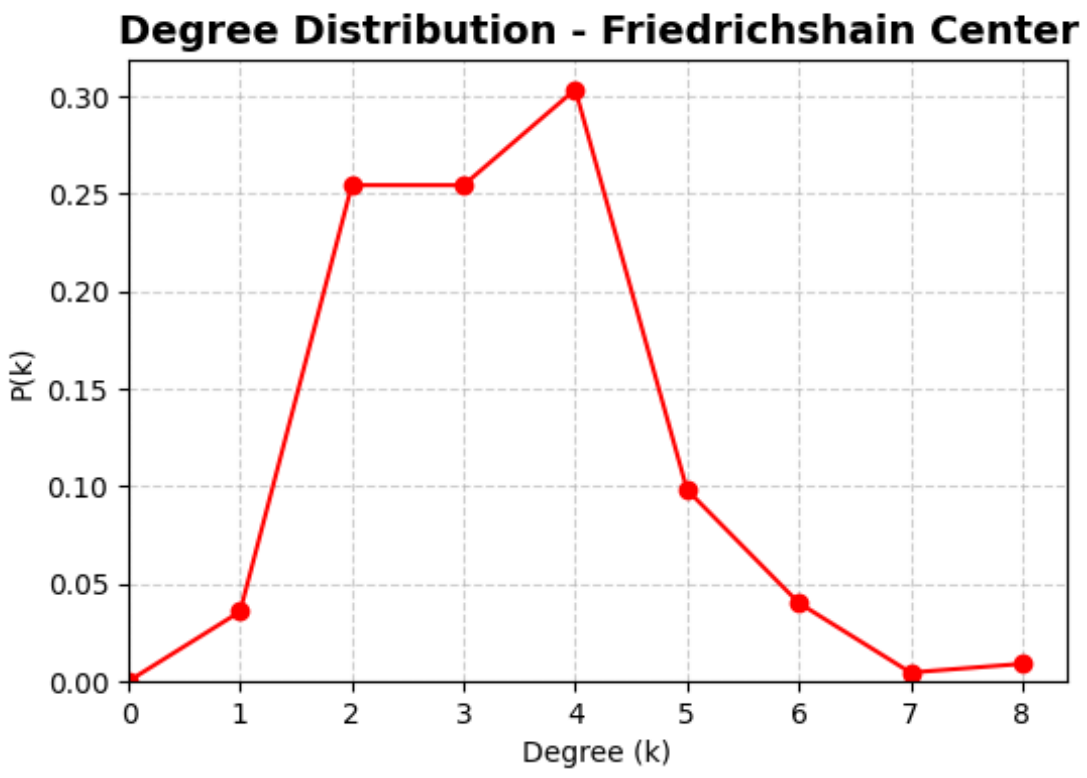

Figure 3. Degree distributions of BFC nodes

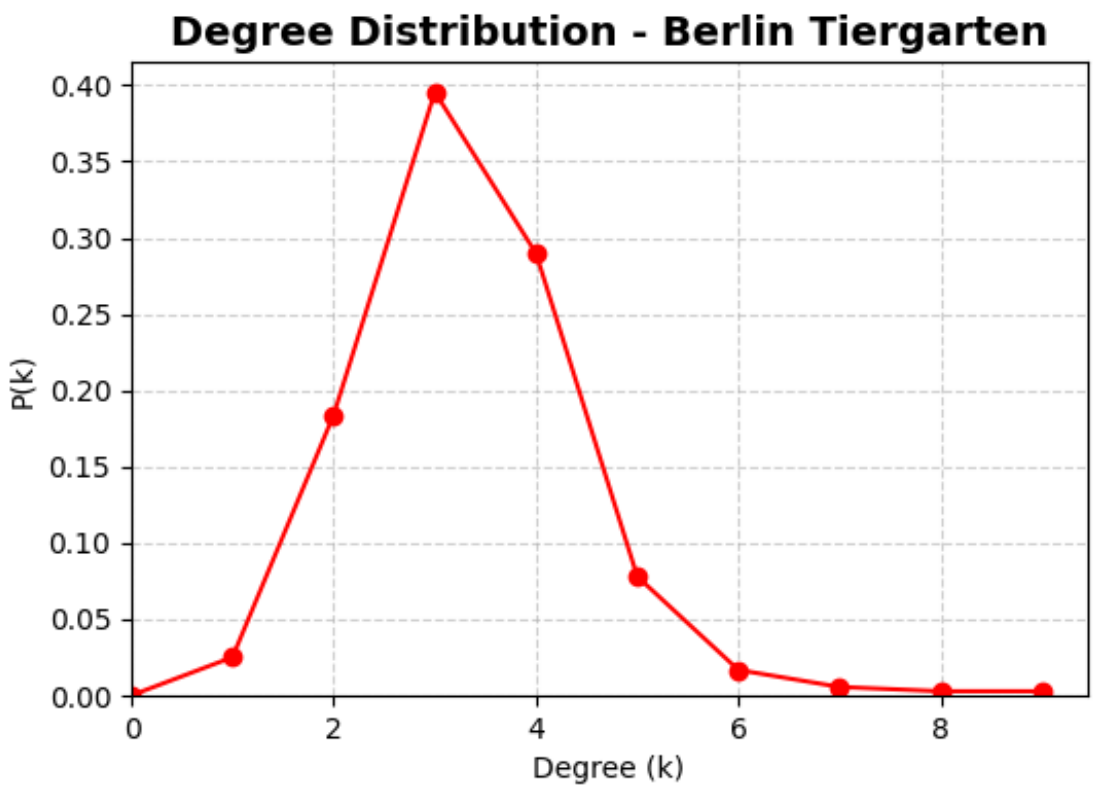

Figure 4. Degree distributions of BT nodes

### 3.3. Global measures based on spectral analysis of the network's graph

Before applying the GW distance, we first establish baseline structural characteristics of our networks through spectral graph theory. Spectral analysis examines the eigenvalue properties of key graph matrices to reveal fundamental structural properties that are not immediately apparent from simple topological measures. This analysis serves two purposes: (1) it provides a comprehensive structural foundation for understanding our networks before edge removal experiments, and (2) it establishes reference points for interpreting how GW distance captures structural changes that complement spectral properties.

The **adjacency matrix spectrum**, characterized by ordered eigenvalues $e_1(G) \leq e_2(G) \leq \cdots e_n(G)$ encodes fundamental structural properties of the network topology. The **spectral radius** of (the adjacency matrix of) a graph $G$ is $e_n(G)$ is denoted by $r(G)$ satisfies the inequality, $\sqrt{\Delta(G)} \leq r(G) \leq \Delta(G)$ establishing a quantitative relationship between local node connectivity and global network properties, where $\Delta(G) = \max_{v \in V(G)} deg(v)$ and deg(v) is the degree of node v [38]. The **spectral gap**, defined as $s(G) = r(G) - e_{n-1}(G)$, serves as a critical vulnerability indicator, where small values denote the presence of structural bottlenecks and bridge components whose removal leads to fragmentation of the network. The Laplacian matrix of G is an $n \times n$ matrix $L(G) = D(G) - A(G)$ constructed from the (diagonal) degree matrix $D(G) = diag(k_i)$ where $k_i$ denotes the degree of the node i and the adjacency matrix $A(G)$, yields non-negative eigenvalues $\lambda_1(G) = 0 \leq \lambda_2(G) \leq \cdots \leq \lambda_n(G)$ that encode connectivity properties. The second smallest Laplacian eigenvalue represents the algebraic connectivity. Algebraic connectivity is one of the most broadly extended measures of connectivity and constitutes a fundamental robustness metric. Larger values of algebraic connectivity represent higher robustness against efforts to disconnect the graph, so the larger it is, the more difficult it is to cut a graph into independent components. The inequalities bound the algebraic connectivity:

$$\lambda^2(G) \leq \frac{n}{(n-1)} \times \min_{k=1:n} \{d_k(G)\} \quad (1)$$

$$\lambda^2(G) \geq \frac{4}{(n \times diameter(G))} \quad (2)$$

Large values of algebraic connectivity correspond to a high resistance against network disconnection.
Our empirical analysis reveals differential vulnerability characteristics through these spectral measures. Specifically, $\lambda_2(BFC) = 0.0209$ and $\lambda_2(BT) = 0.0055$, demonstrating that BT exhibits a particularly low resilience to disruptions. The spectral gap analysis further differentiates network vulnerability: with $s(BFC) = 0.1763$ indicating moderate structural cohesion, while $s(BT) = 0.7745$ suggests a higher structural integrity despite lower algebraic connectivity.
These spectral invariants provide complementary insights to traditional topological measures such as network diameter and degree distribution, enabling a comprehensive assessment of global connectivity patterns and vulnerability to systematic disruptions without requiring computationally intensive attack simulations. While spectral analysis reveals global structural properties of the intact network, it cannot predict the specific geometric reorganization that occurs when individual edges are removed. This limitation motivates our subsequent methodology using the GW distance, which directly quantifies how edge removals reshape the entire network geometry, providing complementary insights to spectral analysis.

## 4. Methods

### 4.1. Gromov-Wasserstein distance

The GW distance is a generalization of the most well-known Wasserstein distance that was proposed in the Optimal Transport (OT) theory to compare probability measures (aka probability distributions). While Wasserstein distance requires that the two probability measures to be compared must be fully supported on the same space – even if their supports can be different sets in this space – the GW distance overcomes this limitation, leading to the possibility to compare two completely different measure metric spaces. Formally, a **measure metric space** (**mm-space**) is a triple $\mathcal{X} = (X, d_X, \mu)$ with $X \subseteq \mathbb{R}^{h_x}$ not empty and a separable and completely metrizable space (i.e., $X$ is a Polish space), and $d_X: X \times X \to \mathbb{R}$ a (distance) metric generating the geometry of the space $X$. The metric space $(X, d_X)$ is a Polish metric space if and only if $X$ is a Polish space and the metric $d_X$ is one of the complete metrics compatible with the topology.
More interesting, GW also applies to the so-called measure network spaces, where a **measure network space** (**mnet-space**) is defined as a measure metric space, but where the distance metric $d_X$ is replaced with a **network function** $\omega_X: X \times X \to \mathbb{R}$, which is not necessarily a distance. Indeed, GW is nowadays one of the key concepts for the analysis of graphs as well as complex data remapped as graphs. It extends optimal transport to cases where distributions cannot be compared point by point [12], [39].

Since we are working with transportation networks, the space $X$ is finite and consists of all the network's nodes, that is $X = V$. The probability measure is just uniform, meaning that $\mu$ is an empirical distribution, namely $\mu = \frac{1}{n}\sum_{i=1:n}\delta_{x_i}$, with $\delta_{x_i}$ the Dirac's delta function centered in $x_i \in X$. The network function $\omega_X(x_i, x_j)$ is, in our study, the shortest path between the nodes $x_i$ and $x_j$. It is important to remark that, although the shortest path is based on the length of the edges, it does not necessarily lead $\omega_X$ to being a distance: the length of an edge between two nodes, in the real network, is not necessarily the Euclidean distance between those nodes. Indeed, the most general notion of network function must be considered.

Finally, the GW distance between two m-net spaces, namely $\mathcal{X} = (X, d_X, \mu)$ and $\mathcal{Y} = (Y, d_Y, \nu)$ is obtained by solving the following quadratic optimization problem:

$$GW_p(\mathcal{X}, \mathcal{Y}) = \min_{T \in \mathcal{U}(\mu,\nu)} \sum_{i,í=1}^{n} \sum_{j,ј=1}^{m} \mathcal{L}(C_X[i, i'], C_Y[j, j'])T_{i,j}T_{i',j'} \tag{3}$$

where $p$ is the order of the GW distance, $\mathcal{U}(\mu, \nu) = \{T \in \mathbb{R}_+^{n\times m} : T1_m = \mu,\ T^\top 1_n = \nu\}$ is the set of feasible solutions consisting of all the possible joint probability measures having $\mu$ and $\nu$ as marginals, and $\mathcal{L}(\cdot,\cdot)$ is a loss function, usually the squared loss – i.e., $\mathcal{L}(a, b) = (a - b)^2$. Finally, $C_X[i, i'] = \omega_X(x_i, x_{i'})^p$ and $C_Y[j, j'] = \omega_Y(y_j y_{j'})^p$.

The coupling $T$ represents a soft assignment between nodes of the two spaces. The GW distance is thus well-suited for comparing structured m-net spaces, where the intrinsic geometry of the objects matters more than their embedding in a common space. This property makes GW particularly appropriate for transport networks.

### 4.2. Comparing original versus damaged transportation network via GW

To the best of our knowledge, the present paper is the first to apply this framework in the context of transport network vulnerability. In our approach, the m-net space $\mathcal{X}$ is the original transport network, while $\mathcal{Y}$ is the same network after the removal of a specific edge (i.e., disruption of a connection in the real-world transportation network). It is easy to understand that $X = Y$ (because nodes are not removed), as well as $\mu = \nu$ (for the same reason). Moreover, $\omega_X = \omega_Y$ because, in the two m-net spaces they are the shortest path. However, the adjacency matrix is different, even if just a single edge has been removed, leading to $C_X \neq C_Y$ and, thus, to $GW_p(\mathcal{X}, \mathcal{Y}) > 0$, quantifying how much the intrinsic geometry of the network has been altered by the removal of that specific edge. In other words, GW provides a principled way to measure the global consequence of edge removals, capturing distortions in the distribution of all-pairs shortest-path distances. This allows us to quantify how much the global geometry of the network changes after disruptions – capturing resilience effects that are not reflected by local centrality measures or maximum shortest path length alone. Figure 5 illustrates the principle of comparing two metric spaces using GW.

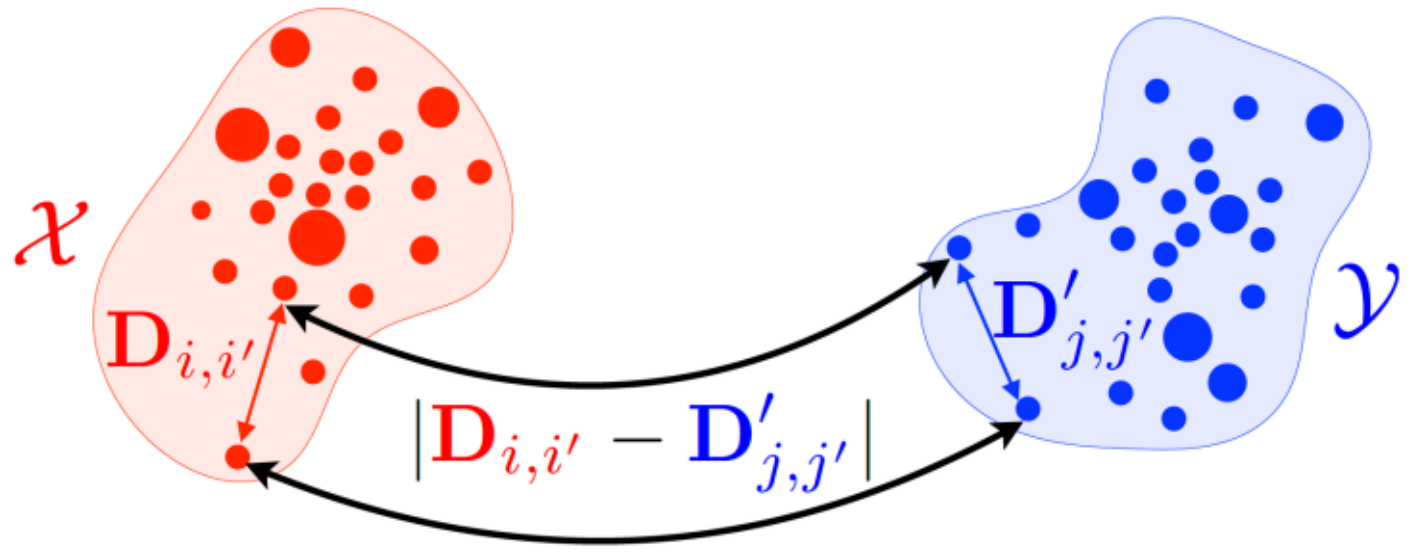

Figure 5. The GW approach to comparing two metric measure spaces [15].

As already mentioned, computing GW is equivalent to solve a quadratic problem, specifically a Quadratic Assignment Problem (QAP) in the case $n = m$, that is exactly our case, because $X = Y$.
Since QAP is NP-hard [40], computing exact GW distances is intractable for large graphs. In practice, approximate methods are employed, such as entropic regularization and iterative linearization schemes [12]. These methods produce stationary points rather than global optima, but they are computationally feasible and effective for real-world networks. Recent work has also introduced histogram-based relaxations and lower bounds [41], which reduce computational cost by approximating global distance distributions and using them as initialization for GW solvers.

## 5. Experimental setup

### 5.1. Protocol

We have designed and adopted an edge-based experimental protocol: for each edge $e_k \in E$, we remove it from the original network, that is the m-net space $\mathcal{X}$, and obtain the new m-net space $\mathcal{Y}_k$, where the subscript $k$ relates to the removed edge $e_k$. According to the two different adjacency matrices, the two cost matrices $C_X$ and $C_{Y_k}$ are computed – again, the subscript $k$ relates to the removed edge $e_k$ – and the associated GW distance is computed, namely $GW_p(\mathcal{X}, \mathcal{Y}_k)$. The larger the value of $GW_p(\mathcal{X}, \mathcal{Y}_k)$ the more relevant the structural change induced by the removal of the edge $e_k$. Ranking all the edges depending on their values of $GW_p(\mathcal{X}, \mathcal{Y}_k)$ allows for identifying the most critical edges with respect to the preservation of the overall network connectivity. Moreover, the optimal transport plan $T^*$ associated to the computation of the GW provides important information by itself: it indicates how node-to-node relationships are reweighted in the perturbed network compared to the baseline. This allows us to capture not only global distortions but also localized redistribution effects across the network. Such transport-based perspectives on network disruption have recently been suggested as a promising complement to classical graph-theoretic measures.

### 5.2. Criticality based on local measure (edge betweenness)

We compare results from our GW-based analysis against the local graph-theoretic measure more directly correlated to the shortest path concept, that is the edge betweenness. Formally, edge betweenness of an edge is defined as the number of shortest paths passing through that edge, considering the shortest paths between every possible pair of nodes in the graph:

$$B(e_k) = \sum_{i \neq j} \frac{\sigma\left(i \overset{e_k}{\rightarrow} j\right)}{\sigma(i \rightarrow j)} \tag{4}$$

where $\sigma\left(i \overset{e_k}{\rightarrow} j\right)$ denotes the number of shortest paths from node $i$ to node $j$ passing through the edge $e_k$, and $\sigma(i \rightarrow j)$ denotes the number of shortest paths from node $i$ to node $j$, overall.

Edges having a high edge betweenness value are usually regarded as structurally important because they mediate flows between many node pairs [42], [43].

### 5.3. Criticality based on a proxy of the service level

Traditionally, metrics such as edge betweenness centrality have been employed to identify critical links, reflecting how frequently an edge participates in shortest paths. While betweenness captures traffic concentration, it does not necessarily measure how the global structure of accessibility is affected by disruption of connections. To address this, we compare the GW distance with a structural indicator based on the maximum shortest path (MSP) of the network (aka diameter). Specifically, we consider the variation in MSP between the network before and after the disruptive event, that is:

$$\Delta_{MSP}(e_k) = \max_{i,j=1:n} \{C_{Y_k}\} - \max_{i,j=1:n} \{C_X\}$$

A large value of $\Delta_{MSP}(e_k)$ indicates that the disruption of the connection $e_k$ significantly elongates the longest travel path, thereby reducing resilience.

We conduct correlation analyses between $GW_p(\mathcal{X}, \mathcal{Y}_k)$, $B(e_k)$, and $\Delta_{MSP}(e_k)$ to evaluate the extent to which GW provides complementary information. This comparative framework allows us to test the hypothesis that GW captures both global geometric distortions (beyond $\Delta_{MSP}(e_k)$) and distributed connectivity effects (beyond $B(e_k)$).

## 6. Results and discussion

The results demonstrate that GW distance provides a consistent and discriminative ranking of edge criticality. In the BFC network, the top-ranked edges by GW distance (Table 2) correspond to links whose removal significantly perturbs the global distribution of shortest paths. For example, edges (171,224) and (88,213) exhibit GW distances above 1.47, an order of magnitude higher than peripheral edges such as (192,196) with a score near 0.0014. This sharp contrast indicates that GW not only separates critical from redundant edges but also assigns meaningful magnitudes to their structural impacts. Figure 8 visualizes the spatial distribution of the highest-GW edges in BFC. The critical edges (highlighted in red) are distributed across different regions of the network rather than being concentrated in a single area. These edges appear to serve as key connectors between different parts of the network topology, and their strategic positions suggest they play important roles in maintaining overall network connectivity. The spatial pattern indicates that the removal of these edges would create significant geometric distortions in the distance matrix by disrupting multiple connection pathways simultaneously, which explains the pronounced impact on global distance distributions measured by GW distance. To ensure reproducibility and facilitate further research, all code used for network analysis, GW distance computation, and visualization is publicly available at https://github.com/iman-ie/gw-net-vuln. The computations were carried out on a machine equipped with an Intel i7-9700 CPU and 16 GB RAM. For GW distance estimation, we employed the Python Optimal Transport (POT) library [44], together with supporting libraries such as NetworkX for graph processing, NumPy/SciPy for numerical routines, and Matplotlib for visualization.

Table 2. Top 10 edges with the highest GW distance in BFC

| Edge | GW distance |
|---|---|
| (171, 224) | 1.476623 |
| (88, 213) | 1.476562 |
| (55, 222) | 1.475641 |
| (131, 132) | 1.475640 |
| (56, 54) | 1.475630 |
| (112, 130) | 1.405670 |
| (182, 212) | 1.404043 |
| (71, 24) | 0.002371 |
| (215, 217) | 0.001702 |
| (192, 196) | 0.001417 |

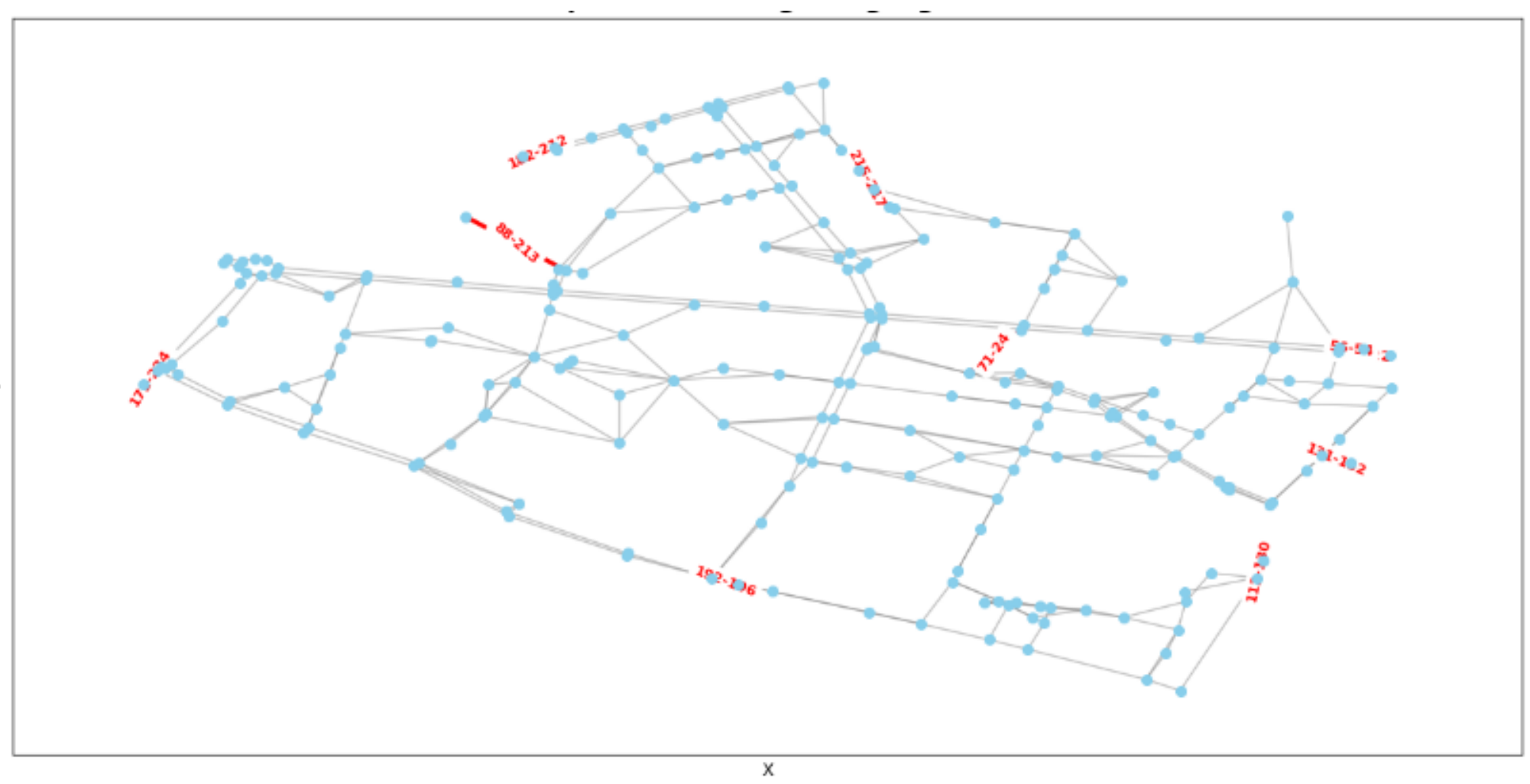

Figure 8. Edges with the highest GW distance, in BFC Network

The same trend is visible in the BT network (Table 3), but with substantially larger values. The most disruptive edge (73,232) reaches a GW distance above 5.6, far exceeding the critical values observed in BFC. As mentioned before in section 3.1, BT's significantly lower algebraic connectivity ($\lambda = 0.089$) indicates weaker overall structural cohesion, making the network more dependent on individual critical connections to maintain global connectivity. When these structurally essential edges are removed, the resulting geometric distortions are amplified precisely because the network lacks robust alternative pathways—a vulnerability encoded in its low algebraic connectivity. Conversely, while BFC's higher algebraic connectivity ($\lambda = 0.127$) suggests better distributed connectivity, individual edge failures produce more contained geometric distortions. Figure 9 shows that these critical edges are concentrated in specific localized regions of the network, forming clusters of vulnerability rather than being distributed along central corridors. The spatial pattern reveals that the most critical edges (highlighted in red) are located in densely connected subregions where their removal would isolate or significantly disconnect these local clusters from the broader network. This clustering pattern indicates that BT's vulnerabilities stem from structural dependencies within specific neighborhoods rather than from dispersed bridging connections. The localized concentration of high-GW edges suggests that certain areas of the BT network act as critical junction points whose failure would have disproportionate impacts on overall connectivity. Compared to BFC, BT demonstrates greater susceptibility to disruptions, as evidenced by both the magnitude of GW distances and the concentrated spatial clustering of critical edges in vulnerable zones.

Table 3. Top 10 edges with the highest GW distance in BT

| Edge | GW distance |
|---|---|
| (73, 232) | 5.632254 |
| (232, 234) | 4.226409 |
| (231, 230) | 3.083483 |
| (76, 77) | 1.545568 |
| (73, 78) | 1.545383 |
| (237, 231) | 1.545374 |
| (113, 109) | 1.545245 |
| (104, 93) | 1.545039 |
| (105, 89) | 1.544971 |
| (252, 254) | 1.544744 |

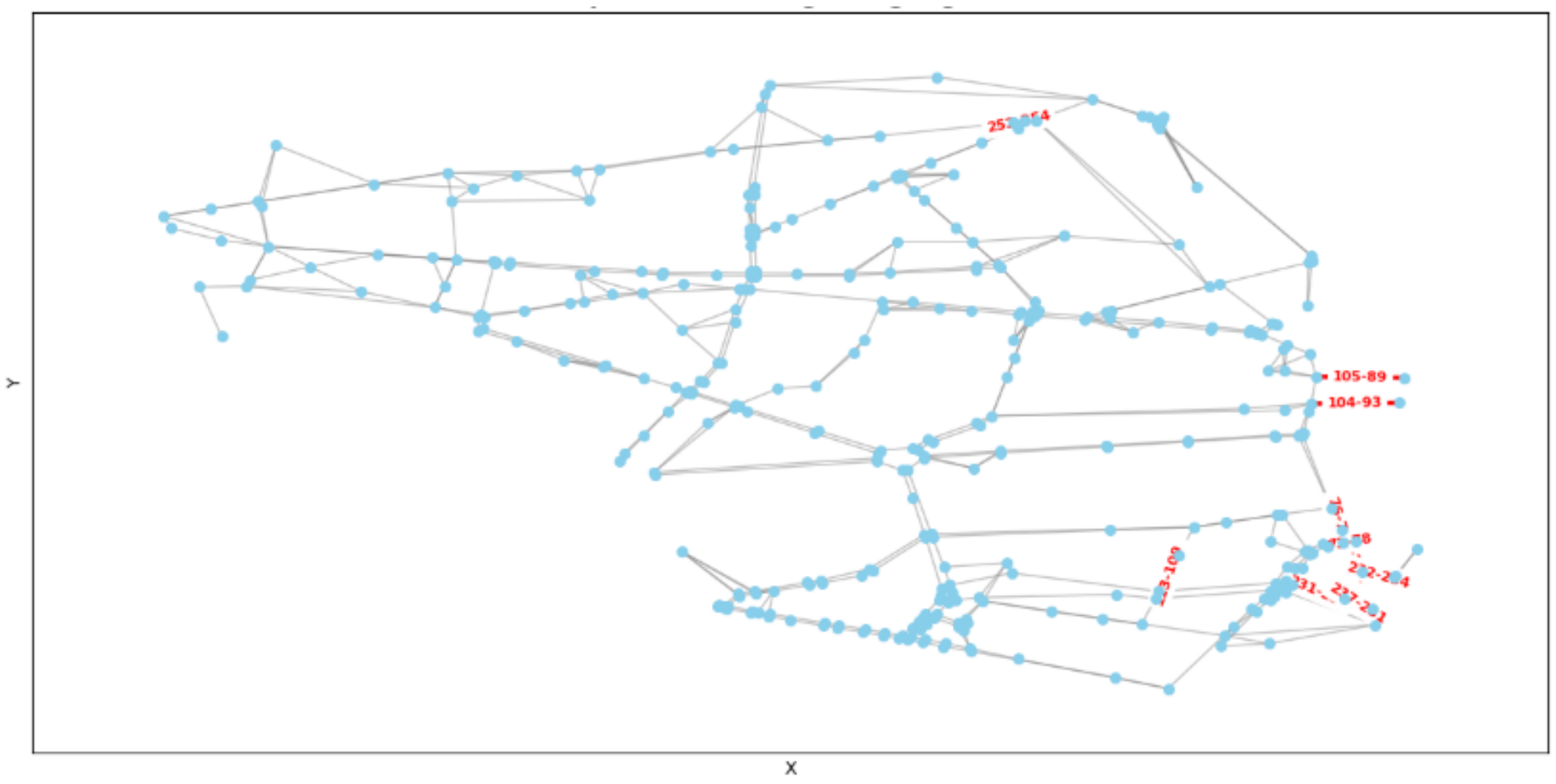

Figure 9. Edges with the highest GW distance, in BT Network

Node-level experiments provide additional insights into the relationship between local network properties and global structural vulnerability. Table 4 examines the impact of removing all edges connected to the highest-degree nodes in each network. In BFC, removing node 201 (degree 8) produces a GW distance of 1.486, while the analogous removal of node 215 (degree 9) in BT yields a substantially higher value of 2.028. This 36% increase in GW distance, despite only a marginal difference in node degree, reinforces the spectral analysis findings that BT's lower algebraic connectivity ($\lambda = 0.089$) makes it more vulnerable to structural perturbations than BFC ( $\lambda = 0.127$).

More significantly, Table 5 reveals that even median-degree nodes (degree 3 in both networks) generate substantial GW impacts, with values of 1.476 in BFC and 1.545 in BT. These results are particularly noteworthy because they approach the magnitude of highest-degree node removals, challenging the conventional assumption that structural criticality correlates directly with node degree. The fact that median-degree nodes produce GW distances within 1-4% of those of the highest-degree nodes demonstrates a fundamental limitation of degree-based vulnerability assessments. Furthermore, the systematic difference between BFC and BT responses (BT consistently showing 4-36% higher GW distances) aligns with the spectral characteristics identified earlier, where BT's structural properties indicated greater susceptibility to fragmentation. These node-level results thus corroborate the edge-level findings, demonstrating that GW distance captures structural criticality that extends beyond traditional centrality-based predictions. This reveals hidden fragilities in network geometry that degree-based assessments would systematically underestimate.

Table 4. GW distance after removing all edges of the Highest degree node

| Network | Node (degree) | Edges | GW distance |
|---|---|---|---|
| BFC | 201 (8) | (201,7)-(201,21)-(201,200)-(201,202)-(201,205)-(201,207)-(201,208)-(201,210) | 1.486825 |
| BT | 215 (9) | (215,20)-(215,204)-(215,209)-(215,212)-(215,213)-(215,214)-(215,217)-(215,218)-(215,326) | 2.027595 |

Table 5. GW distance after removing all edges of the Median degree node

| Network | Node (degree) | Edges | GW distance |
|---|---|---|---|
| BFC | 161 (3) | (161,1)-(161,32)-(161,159) | 1.476204 |
| BT | 269 (3) | (269,1)-(269,226)-(269,348) | 1.545100 |

The comparison with classical metrics provides further insight. These relationships are summarized in Table 6, which compares the correlations of GW distance with classical metrics across both networks. In the BFC network, the correlation between GW distance and edge betweenness is effectively zero ($\rho = -0.0542$). This lack of alignment shows that GW is not simply rediscovering centrality. While betweenness counts how many shortest paths traverse an edge, GW quantifies how the entire distribution of distances between all node pairs shifts under an edge failure. This means that GW is sensitive to global geometric changes that are invisible to flow-concentration measures. The same pattern holds for BT, where the correlation between GW and betweenness remains negligible ($\rho = -0.0334$), reinforcing the conclusion that GW captures orthogonal information. Figures 10 and 11 illustrate that GW distance maintains a low correlation with edge betweenness in both networks.

Table 6. Correlation between $GW_p(\mathcal{X}, \mathcal{Y}_k)$, $B(e_k)$, and $\Delta_{MSP}(e_k)$, in BFC and BT networks, separately.

| **Network** | **Metric Pair** | **Correlation ($\rho$)** |
|---|---|---|
| **BFC** | $GW_p(\mathcal{X}, \mathcal{Y}_k)$ vs $B(e_k)$ | −0.0542 |
| | $B(e_k)$ vs $\Delta_{MSP}(e_k)$ | −0.0543 |
| | $GW_p(\mathcal{X}, \mathcal{Y}_k)$ vs $\Delta_{MSP}(e_k)$ | 0.9999 |
| **BT** | $GW_p(\mathcal{X}, \mathcal{Y}_k)$ vs $B(e_k)$ | −0.0334 |
| | $B(e_k)$ vs $\Delta_{MSP}(e_k)$ | −0.0513 |
| | $GW_p(\mathcal{X}, \mathcal{Y}_k)$ vs $\Delta_{MSP}(e_k)$ | 0.8530 |

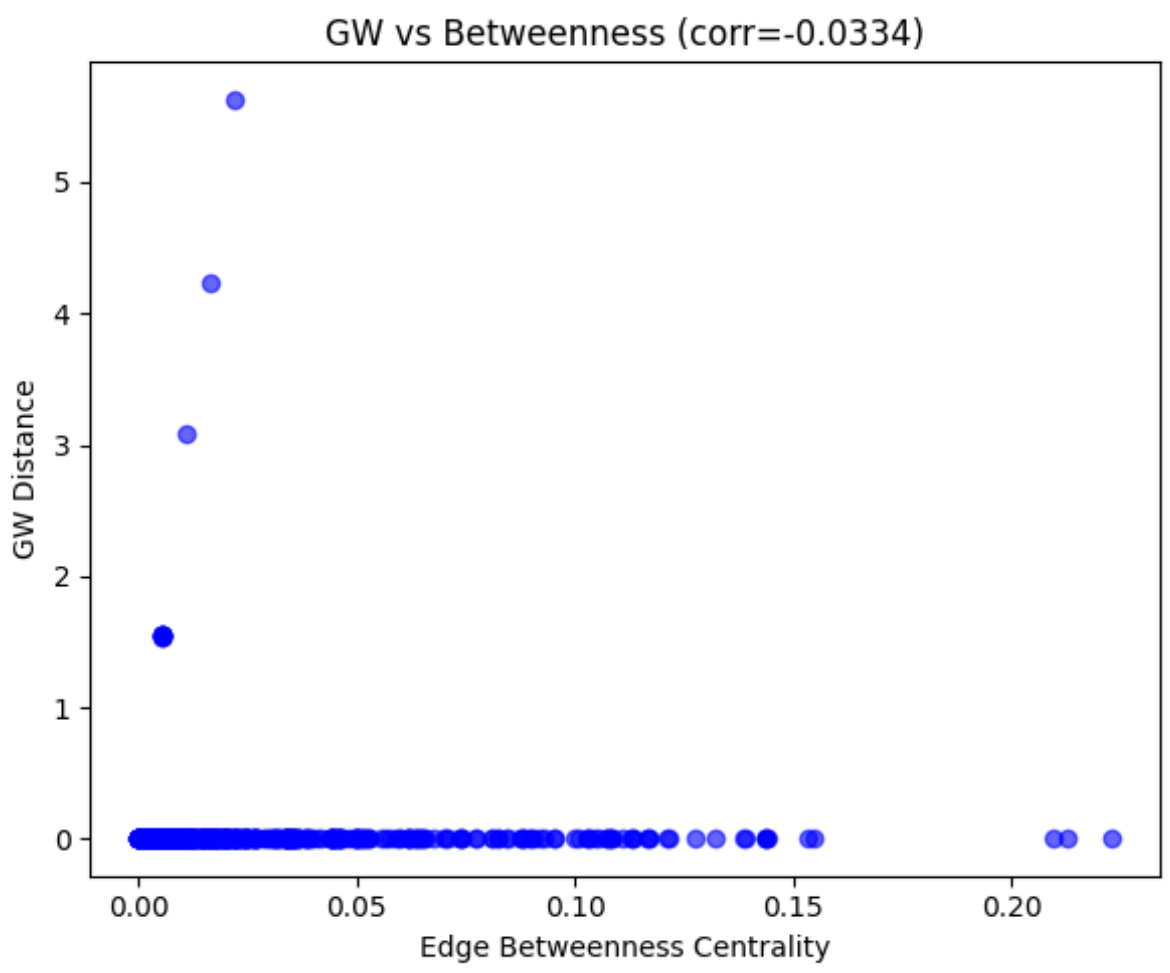

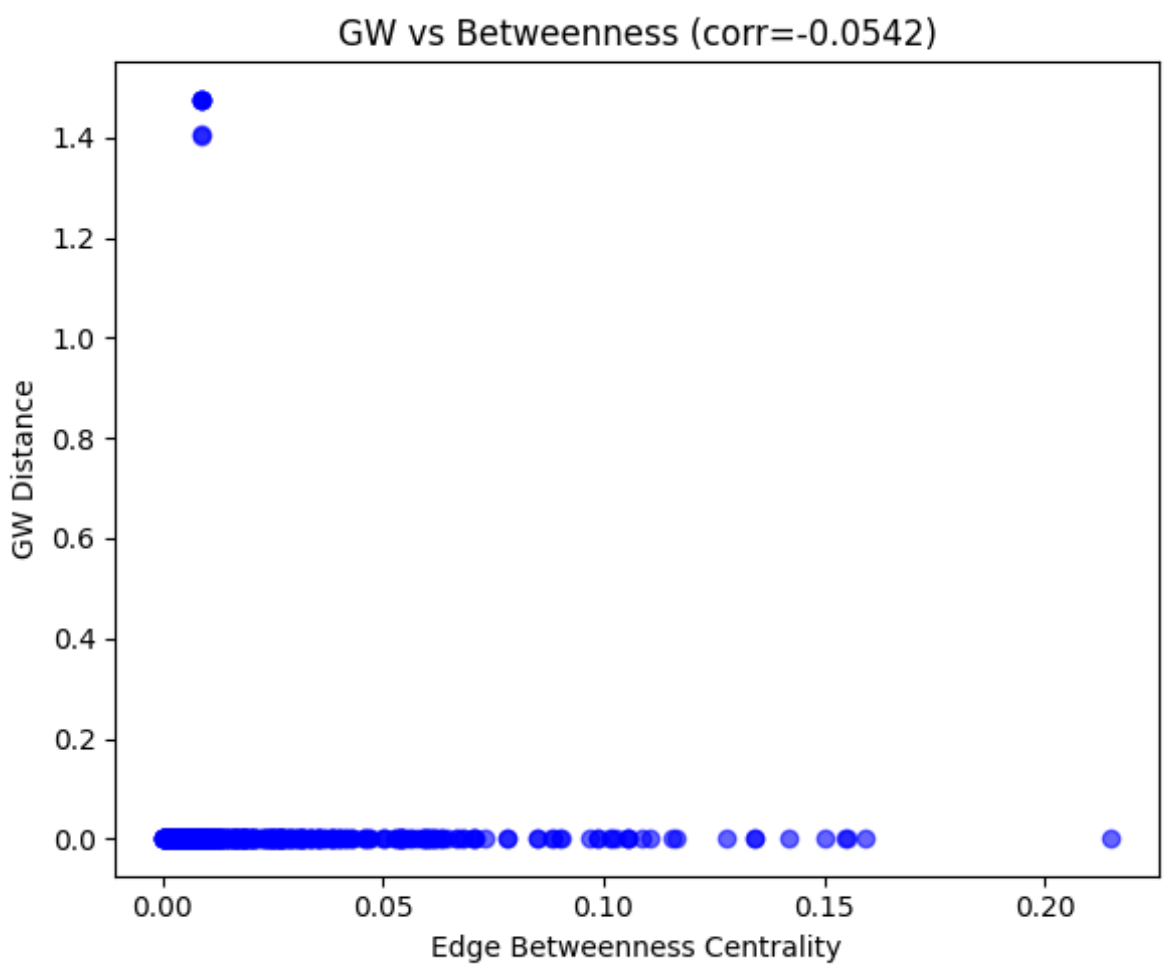

Figure 10. Correlation between GW distance and betweenness, in BT Network

Figure 11. Correlation between GW distance and betweenness, in BFC Network

By contrast, the correlation between GW and the $\Delta_{MSP}$ is almost perfect. In BFC, the correlation reaches ($\rho = 0.9999$), showing that GW internalizes the same global stretch information that $\Delta_{MSP}$ is designed to capture. For BT, the value is comparably strong ($\rho = 0.8530$), further confirming the consistency of this relationship. In BFC, the correlation between betweenness and $\Delta_{MSP}$ is similarly weak ($\rho = -0.0543$), suggesting that betweenness centrality fails to capture the distributed structural impact of edge removals. For BT, the correlation between betweenness and $\Delta_{MSP}$ is also negligible ($\rho = -0.0513$). Figures 12 and 13 illustrate that GW distance maintains an extremely strong positive correlation with $\Delta_{MSP}$ in both networks.

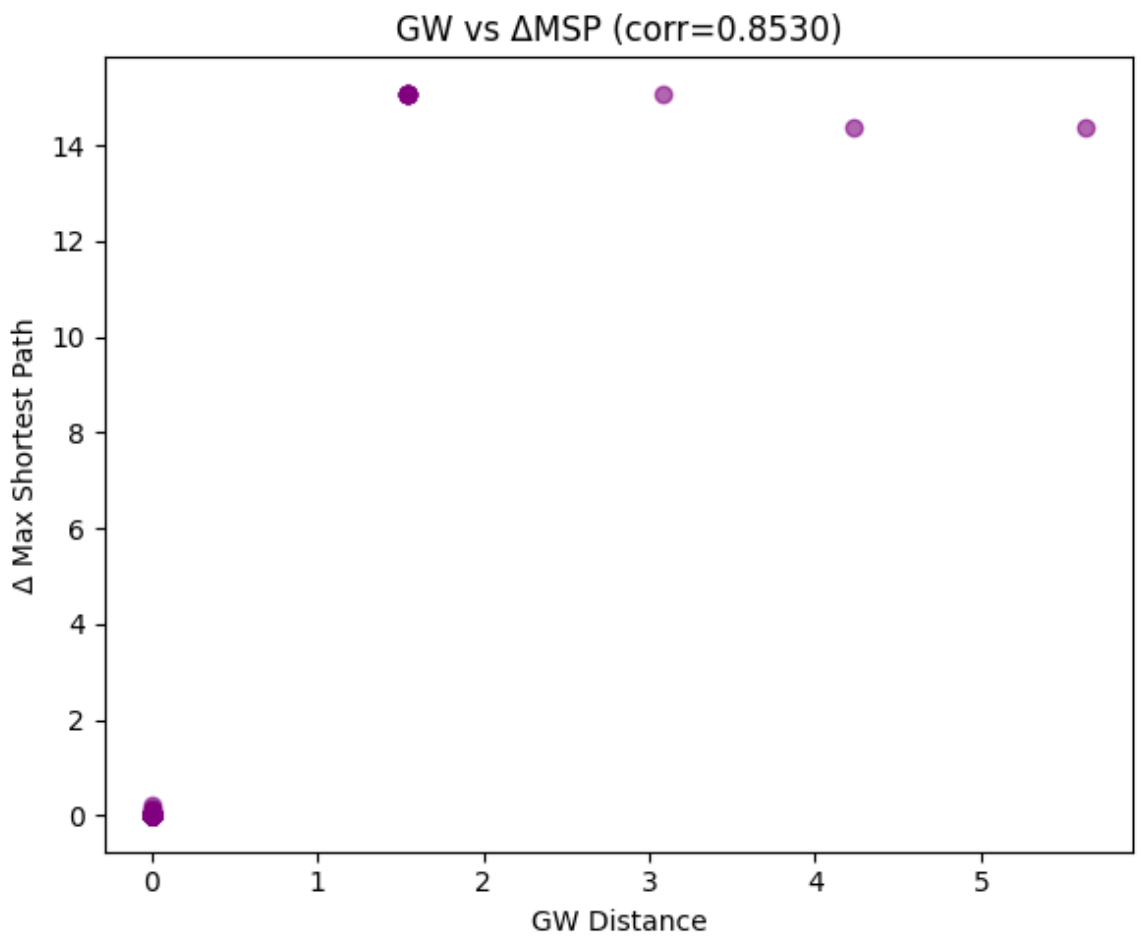

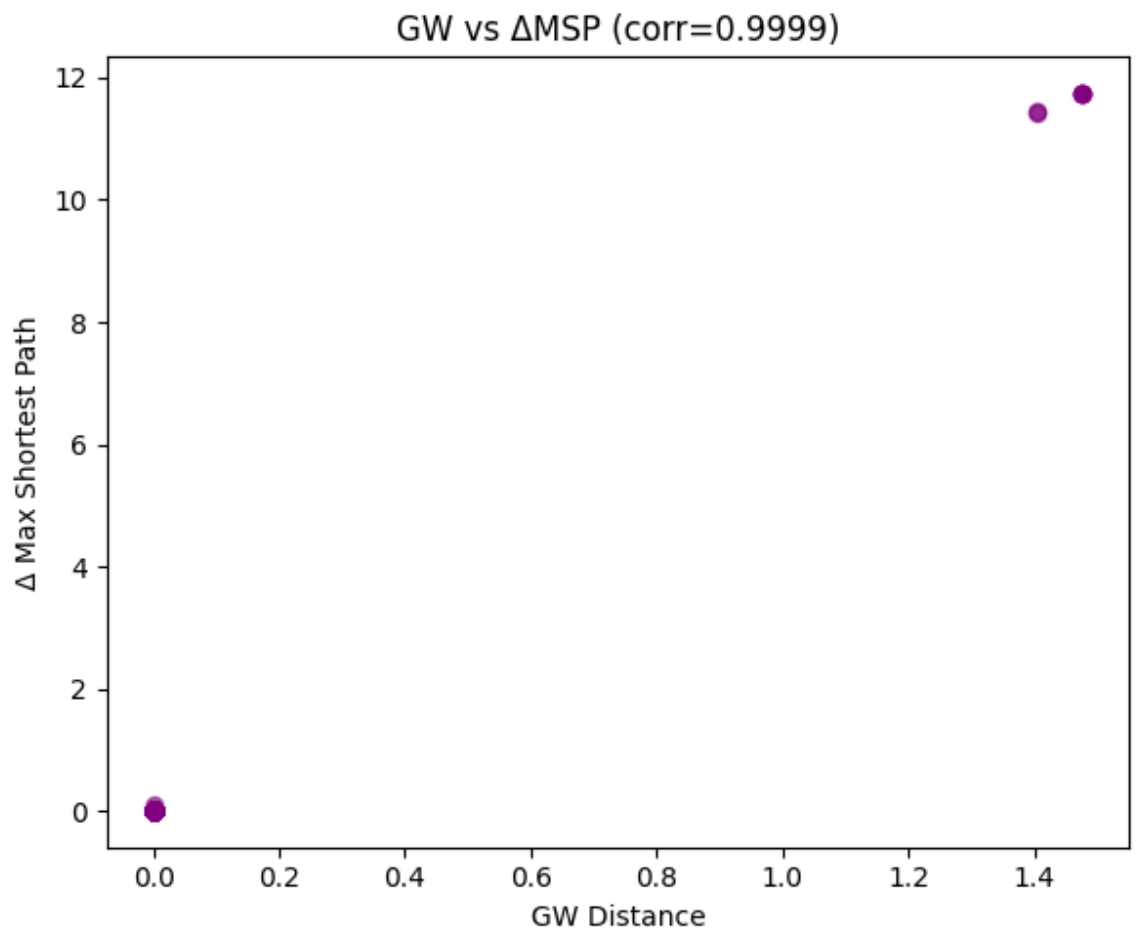

Figure 12. Correlation between GW distance and MSP, in BT Network

Figure 13. Correlation between GW distance and MSP, in BFC Network

In the BT network, there are 516 edges whose removal leads to GW distance close to zero (i.e., GW<0.1), while 7 edges are associated to a GW>1.4, without any other edge in between of these two ranges. Similarly, the BFC network there are 754 edges whose removal leads to GW distance close to zero and 12 edges with an associated GW larger than 10. This pronounced separation indicates two fundamentally different edge characteristics within both network architectures, suggesting that the relationship between geometric-topological distances and maximum shortest path changes follows a distinct binary pattern rather than a continuous distribution.
Results highlight the unique value of the GW distance: betweenness centrality and other local or flow-based measures fail to explain the global changes in accessibility observed after disruptions, while GW accounts for the global reorganization of the flows, measure in terms of shortest paths. This makes GW a global graph metric capable of identifying critical edges and diagnosing vulnerabilities.

## 6. Discussion and Conclusion

This study demonstrates the effectiveness of the Gromov–Wasserstein (GW) distance as a comprehensive metric for quantifying structural vulnerability in urban transportation networks. The empirical analysis reveals that GW distance exhibits sensitivity, capturing both extreme connectivity deterioration ($r \approx 0.9999$ with ΔMSP) and distributed geometric distortions across the network topology. This dual capability distinguishes GW from traditional graph-theoretic measures, which typically focus on either local structural properties or individual path-based characteristics. The negligible correlation between GW distance and edge betweenness centrality ($\rho < 0.1$) demonstrates that flow-based centrality measures systematically fail to identify edges that are critical from a global geometric perspective. This finding suggests that conventional centrality-based approaches may misclassify edge importance, potentially leading to suboptimal infrastructure prioritization strategies. The binary clustering pattern observed in the GW-$\Delta_{MSP}$ correlation space indicates that network edges exhibit fundamentally distinct vulnerability characteristics rather than continuous vulnerability gradients.
From a practical standpoint, GW-based vulnerability assessment can inform targeted maintenance strategies, emergency response protocols, and network redundancy planning by identifying structurally critical edges that classical metrics overlook. However, computational complexity remains a limitation for very large networks, and the current framework does not account for traffic loads or capacity constraints.
Future research should incorporate weighted demand patterns, model cascading failures, and develop computationally efficient approximation algorithms for metropolitan-scale applications. Comparative analysis across diverse urban contexts would further strengthen the generalizability of this approach.
In conclusion, the GW distance provides a theoretically sound, empirically validated framework for identifying structurally critical infrastructure elements, offering urban planners a more comprehensive foundation for resilience-informed decision-making than classical measures alone.